\documentclass[12pt,preprint]{aastex}
\newcommand{\U}{\mbox{$u^*$}}
\newcommand{\G}{\mbox{$g'$}}
\newcommand{\R}{\mbox{$r'$}}
\newcommand{\I}{\mbox{$i'$}}
\newcommand{\Z}{\mbox{$z'$}}
\newcommand{\swarp}{\mbox{\tt SWarp}}

\shorttitle{MegaPipe}
\shortauthors{Gwyn}

\begin{document}


\title{MegaPipe: the MegaCam image stacking pipeline at the Canadian Astronomical Data Centre}


\author{Stephen. D. J. Gwyn}
\affil{
Canadian Astronomical Data Centre,
Herzberg Institute of Astrophysics,
5071 West Saanich Road,
Victoria, British Columbia,
Canada   V9E 2E7
}
\email{Stephen.Gwyn@nrc-cnrc.gc.ca}



\begin{abstract}
This paper describes the MegaPipe image processing pipeline at the
Canadian Astronomical Data Centre. The pipeline combines multiple
images from the MegaCam mosaic camera on CFHT and combines them into a
single output image.  MegaPipe takes as input detrended MegaCam
images and does a careful astrometric and photometric calibration on
them.  The calibrated images are then resampled and combined into
image stacks.  The astrometric calibration of the output images is
accurate to within 0.15 arcseconds relative to external reference
frames and 0.04 arcseconds internally.  The photometric calibration is
good to within 0.03 magnitudes.  The stacked images and catalogues
derived from these images are available through the CADC website.
\end{abstract}




\section{INTRODUCTION}

The biggest barrier to using archival MegaCam \markcite{megacam}({Boulade} {et~al.} 2003) images
is the effort required to process them. While the individual images
are occasionally useful by themselves, more often the original
scientific program called for multiple exposures on the same field in
order to build up depth and get rid of image defects. Therefore, the
images must be combined.

A typical program calls for 5 or more exposures on a single
field. Each MegaCam image is about 0.7Gb (in 16-bit integer format),
making image retrieval over the web tedious. Because of the distortion
of the MegaPrime focal plane, the images must be resampled. This
involves substantial computational demands. During this processing,
which is often done in a 32-bit format, copies of the images must be
made, increasing the disk usage. In summary, the demands in terms of
CPU and storage are non-trivial.  Presumably Moore's law (1965),
\markcite{moore} will make these concerns negligible, if not laughable,
in ten years time.  However, at the moment, they present a
technological barrier to easy use of MegaCam data.

The Elixir pipeline \markcite{elixir}({Magnier} \& {Cuillandre} 2004) at CFHT processes each MegaCam
image. It does a good job of detrending (bias-subtracting,
flat-fielding, de-darking) the images. However, the astrometric
solution Elixir provides is only good to 0.5-1.0 arcseconds. To combine
the images, they must be aligned to better than a pixel. One arcsecond
accuracy is insufficient. Therefore, it is necessary for a user to
devise some way of aligning the images to higher accuracy. This is not an
easy task, and rendered more difficult by the distortion of the
MegaPrime focal plane. The problem is not intractable and there do
exist a number of software solutions to the problem, but it remains an
obstacle to easy use of MegaCam data.

In short, while the barriers to using archival MegaCam data are not
insurmountable, they make using these data considerably less
attractive. MegaPipe aims to increase usage of MegaCam data by
removing these barriers.  This paper describes the MegaPipe image
processing pipeline.  MegaPipe combines MegaCam images into stacks and
extracts catalogues.

The procedure can be broken down into the following steps:
\begin{itemize}
\item Image grouping (Section \ref{grouping})
\item Astrometric calibration (Section \ref{astrom})
\item Photometric calibration (Section \ref{photom})
\item Image stacking (Section \ref{comb})
\item Catalogue generation (Section \ref{cat})
\end{itemize}
Sections \ref{qualastro} and \ref{qualphoto} discuss checks on the
astrometric and photometric properties of the output images. Section
\ref{sec:prod} describes the production and distribution of the
images.

\section{IMAGE GROUPING}
\label{grouping}

The first step of image grouping is to ensure that each input image
can be calibrated.

In order for astrometric calibration to take place, each CCD of each
image must contain a minimum number of distinct sources with well
defined centres.  This means that images with short exposure times
($<$50 seconds), images of nebulae or images taken under conditions of
poor transparency cannot be used.

In order for photometric calibration to take place, the image must
either be taken on a photometric night or contain photometric
standards. Only \U\G\R\I\Z\ images are included in MegaPipe, so the
SDSS \markcite{sdss}({York} {et~al.} 2000), which covers a significant fraction of the sky can
be used as a source of photometric standards. Also, sources from previously
processed MegaPipe images can be used as standards.

One CCD of each exposure is inspected visually. Exposures
with obviously asymmetric PSFs (due to loss of tracking) or other
major defects such as unusually bad seeing, bad focus, or poor transparency
are discarded.
Only a sub-raster of one CCD is examined. However, these defects
will affect the entire mosaic, so this examination is sufficient.
The visual examination takes a trained operator about 1 second
per image. The step is semi-automated using an asynchrous
application which retrieves images from the archive before
the are required so there is no delay in loading them.
The time between images is a few milliseconds.

In some of the exposures, one or several of the CCD's
in the mosaic are dead. Other exposures are unusuable
due to being completely saturated. These cases are
detected automatically by examining the statistics
of the pixel values of a subraster of each CCD of each
mosaic. Images that fail these tests are discarded.

Images which pass quality control are then grouped.  Initially, each
image is in a single group containing only itself.  The grouping
algorithm first sorts all the groups by RA.  It then loops through the
groups looking for neighboring groups. Neighbors are defined as a pair
of groups whose centers are within 0.1 degrees of each other.  When
two neighboring groups are found, they are merged into a new single
group.  The center of the new group is set to the average of the
centers of the two original groups, weighted by the number of images
in each group. This process is repeated until no more new neighbors are
found. It converges after 3 or 4 iterations. In principle, because the
groups are sorted by RA and because the group centers shift as groups
merge, group membership determined by this algorithm could be
sensitive to the details of the image centers.  In practice, because
of the way MegaCam is typically used (relatively small dithers, then a
1 degree or more shift to a new field), the algorithm is remarkably
stable. Even doubling the definition of neighbor from 0.1 degrees to
0.2 degrees only adds a small number of groups.

During the grouping described above, no attention is paid to the
filters in which the images were taken; this is done in a second
step. Each group is examined. If it does not contain five or more
images taken in a particular (\U\G\R\I\Z) filter, all the images taken
in that filter are deleted from the group.
The lower limit of four epxosures represents a compromise. Stacking a
smaller number of images means that image defects are not always
perfectly rejected. Setting a larger limit means less data can
processed. Four was chosen so as to include the large amount of
imaging data taken as part of the CFHTLS Very Wide survey which has a
four exposure observing strategy.

\section{ASTROMETRIC CALIBRATION}
\label{astrom}

The first step in the astrometric calibration pipeline is to run the
well known
SExtractor \markcite{hihi}({Bertin} \& {Arnouts} 1996) source detection software
on each image. The parameters are set so as to
extract only the most reliable objects. The detection criteria are set
to flux levels 5 sigma above the sky noise in at least 5 contiguous
pixels.  This catalogue is further cleaned of cosmic rays (by removing
all sources whose half-light radius is smaller than the stellar locus)
and extended objects (by removing sources whose Kron magnitudes are
not within 2 mags of their aperture magnitudes).  This leaves only
real objects with well defined centres: stars and (to some degree)
compact galaxies.

This observed catalogue is matched to the USNO A2.0 astrometric
reference catalogue \markcite{usno}({Monet} \& {et al.} 1998).  The (x,y) coordinates of the
observed catalogue are converted to (RA, Dec) using the initial Elixir
World Coordinate System (WCS). The catalogues are shifted in RA and
Dec with respect to one another until the best match between the two
catalogues is found. 
The number of individual matches between objects in the 
catalogues is maximized as function of RA and Dec offsets.
An object in one catalogue is deemed to match on object in the other
catalogue if the are within a certain radius (intially 2 arcseconds,
shrinking to 0.5 arcseconds as the WCS is refined) of each other and
there are no objects within a second radius (initally 4 arcseconds,
shrinking to 1 arcsecond. This second criteria protects against false
matches from neighboring objects.

Because there are a number of objects in the two catalogues, a small
number of spurious matches can occur. If the number of individual
matches even for the best offset for a particular CCD is low (either
less than 10 or less than half the average of the other CCDs), then
this CCD is flagged as having failed the initial match. This can occur
for example when the initial WCS is unusually erroneous.  In this
case, the WCS for that CCD is replaced with a default WCS and the
matching procedure is restarted.  Once the matching is complete, the
astrometric fitting can begin. Typically 20 to 50 sources per CCD are
found with this initial matching.  As the accuracy of the WCS
improves, the observed and reference catalogues are compared again to
increase the number of matching sources. A larger number of matching
sources makes the astrometric solution more robust against possible
errors (proper motions, spurious detections, etc.) in either the
observed catalogue or the reference catalogue.

The WCS is split into two parts: the linear part, which describes the
tilt, rotation or offset of each CCD with respect to the focal plane;
and the higher order part which describes the distortion of the focal
plane itself. The linear part is represented in a FITS header by the
{\tt CRVALn}, {\tt CRPIXn} and {\tt CDn\underline{~}n} keywords.
The higher order part of the WCS has no standard representation
as of yet  but the most common is the {\tt PVn\underline{~}n} keywords
of \markcite{wcs3}Calabretta {et~al.} (2004). 

The higher order terms are determined on the scale of the entire
mosaic; that is to say, the distortion of the entire focal plan is
measured. This distortion is well described by a polynomial with
second and fourth order terms in radius, $r$, measured from the centre
of the mosaic. The distortion appears to be quite stable over time,
even when one of the lenses in the MegaPrime optics was
flipped\footnote{\url{http://www.cfht.hawaii.edu/Science/CFHTLS-DATA/megaprimehistory.html}}. Determining
the distortion in this way means that only 2 parameters need to be
determined (the coefficients of $r^2$ and $r^4$) with typically (20-50
stars per chip) $\times$ (36 chips) $\cong$ 1000 observations. If the
analysis is done chip-by-chip, a third order solution requires 20
parameters per chip $\times$ 36 chips $=$ 720 parameters. This is
obviously less satisfactory.

In all of this, the object is minimize the size of the average
astrometric residuals which is very complicated function of the
coefficients of the polynomials. The minimization is done
using the non-linear least-squares approach described
by Press {et~al.} (1992) \markcite{NR} modifed to use singular
value decomposition instead of Gauss-Jordan
elimination.

From the global distortion, the distortion local to each CCD is
determined.  The 2-parameter, fourth-order distortion described in
terms of $r$ (the radius to the image centre) is translated into a
10-parameter, third-order distortion described in terms of $x$ and $y$
(the axes of the CCD). Although the number of parameters needed to
describe the distortion for the whole mosaic increases from 2 to 720,
the 720 parameters depend directly and uniquely on the 2 parameter
global radial distortion; only the representation is changed.  The
translation is done in order to be able to represent the distortion with
the {\tt PVn\underline{~}n} keywords. The error introduced by this
translation is less than 0.001 arcseconds.  The linear terms of the
WCS are determined for each CCD individually, after the focal plane
distortion has been removed.

If the group lies within the SDSS, this process is repeated with
the SDSS replacing the USNO as the reference catalogue. The SDSS has a higher
source density and much greater astrometric precision than the USNO.
After the initial matching of the observed catalogues to the external
reference catalogue (USNO or SDSS), the astrometry is further improved
by matching against an internal catalogue generated from MegaPrime
images. 

For the first band of a group to be reduced (the \I-band, if it
exists, otherwise the order of preference is \R, \G, \Z, \U), the
internal catalogue is generated as follows: The observed source
catalogues for each image are cross-referenced to identify sources
common to two or more of the images. These sources are placed in a
master catalogue. The positions of the sources in this master
catalogue are the average of their positions in each of the original
catalogues in which they appeared.  This master catalogue is superior
to the external reference catalogue because it has a higher source
density since the typical MegaCam image goes deeper than either the
USNO or the SDSS.  Further, the positions in the master catalogue are
no less accurate than this external reference catalogue since each
input catalogue was calibrated directly against it.

For the other bands in the group, the image catalogues are first
matched to the USNO to provide a rough WCS and then matched to the
catalogue generated using the first image so as to precisely register
the different bands. The final astrometric calibration has an internal
uncertainty of about 0.04 arcseconds and an external uncertainty of
about 0.2 arcseconds, as discussed in section \ref{qualastro}.

\section{PHOTOMETRIC CALIBRATION}
\label{photom}

\begin{figure}
\plotone{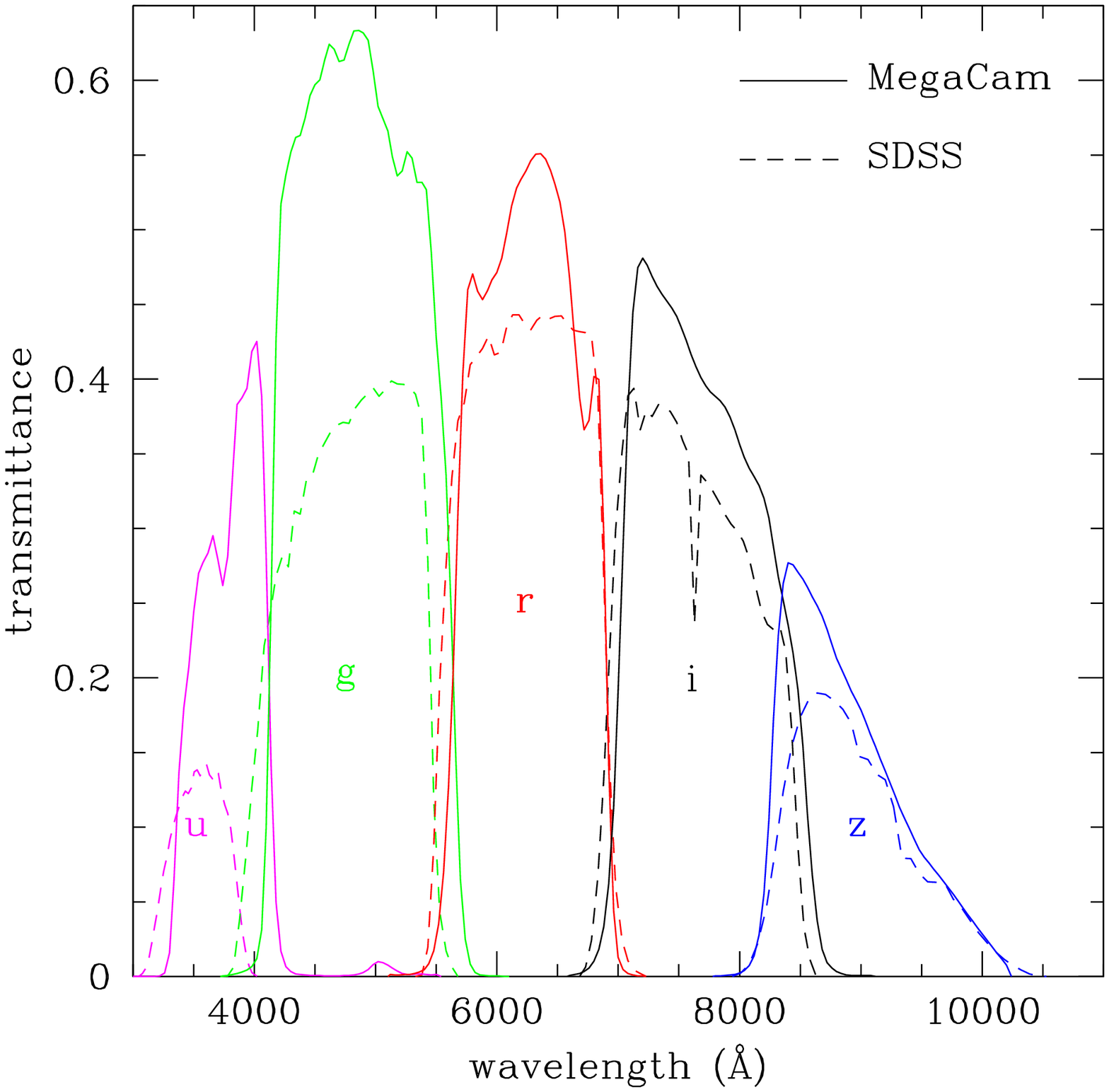}
\caption{Comparing the MegaCam (solid lines) and SDSS (dashed lines) \U\G\R\I\Z filter sets. The transmittance curves
show the final throughput including the filters, the optics and the CCD response.
}
\label{fig:megasdss}
\end{figure}

The Sloan Digital Sky Survey DR5 \markcite{sdssdr5}(Adelman-McCarthy 2007) serves as the basis of the
photometric calibration. The Sloan \U\G\R\I\Z\ filters are not identical to
the MegaCam filters, as shown in Figure \ref{fig:megasdss}.\footnote{see also \url{http://www.cadc.hia.nrc.gc.ca/megapipe/docs/filters.html}}.
The MegaCam filters differ from SDSS filters mostly to adapt to the 
different type of CCD used in the camera.
 The color terms between the
two filter sets can be described by the following equations:

\begin{eqnarray}
\label{eqn:trans1}
\U&=&u_{\rm Mega} = u_{\rm SDSS} - 0.241 (u_{\rm SDSS} - g_{\rm SDSS})\\
\G&=&g_{\rm Mega} = g_{\rm SDSS} - 0.153 (g_{\rm SDSS} - r_{\rm SDSS})\\
\R&=&r_{\rm Mega} = r_{\rm SDSS} - 0.024 (g_{\rm SDSS} - r_{\rm SDSS})\\
\I&=&i_{\rm Mega} = i_{\rm SDSS} - 0.085 (r_{\rm SDSS} - i_{\rm SDSS})\\
\label{eqn:trans5}
\Z&=&z_{\rm Mega} = z_{\rm SDSS} + 0.074 (i_{\rm SDSS} - z_{\rm SDSS}).
\end{eqnarray}

\begin{figure}
\plotone{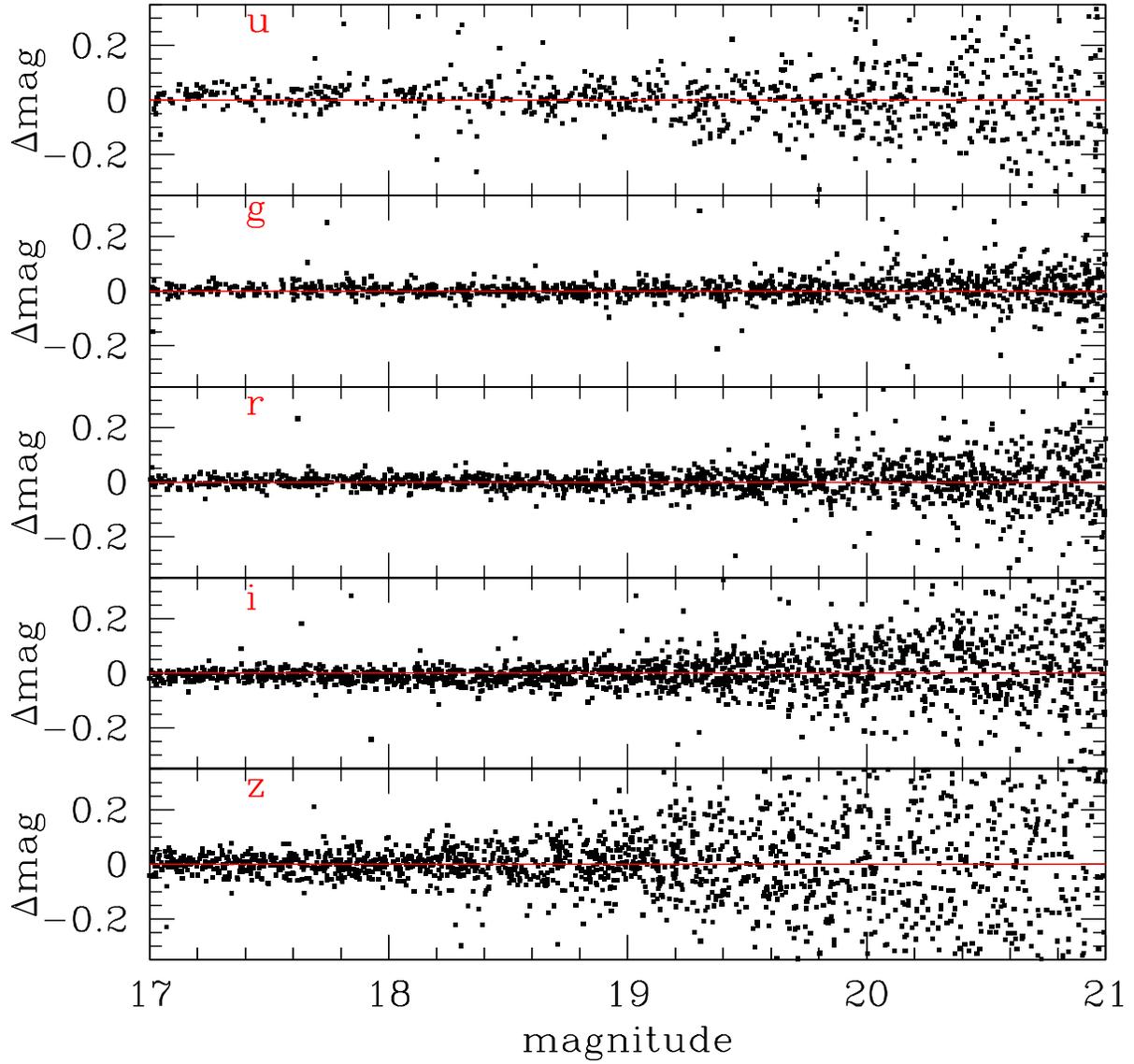}
\caption{The residuals in the transformation between the SDSS and MegaCam \U\G\R\I\Z filter sets.}
\label{fig:sdssmegares}
\end{figure}

The relations for the \G\R\I\Z\ bands come from the analysis of the
SNLS group (Pritchet, private communication)\footnote{\url{http://www.astro.uvic.ca/$\sim$pritchet/SN/Calib/ColorTerms$-$2006Jun19/index.html\#Sec04}}. 
The relation for the \U\ band comes from the CFHT web
pages\footnote{\url{http://cfht.hawaii.edu/Instruments/Imaging/MegaPrime/generalinformation.html}}.
The residuals about these relations are shown in Figure \ref{fig:sdssmegares}.
The amplitude of the residuals in magnitudes where they are not affected by
the intrinsic noise of the photometry are 
$\sigma_u=0.07$, 
$\sigma_g=0.02$, 
$\sigma_r=0.06$, 
$\sigma_i=0.03$, and 
$\sigma_z=0.07$.

All groups lying in the SDSS can be directly calibrated without
referring to other standard stars such as the \markcite{smith}{Smith} {et~al.} (2002) standards.
The systematics in the SDSS photometry are about 0.02 magnitudes
\markcite{ivezic2004}({Ivezi{\'c}} {et~al.} 2004). The presence of at least 1000 usable sources in
each square degree reduces the random error to effectively zero. It is
possible to calibrate the individual CCDs of the mosaic individually
with about 30 standards in each. For each MegaCam image, MegaPipe
matches the corresponding catalogue to the SDSS catalogue for that patch
of sky. The difference between the instrumental MegaCam magnitudes and
the SDSS magnitudes (transferred to MegaCam system using the equations
above) gives the zero-point for that exposure or that CCD. The
zero-point is determined by a median, not a mean. There are about
10000 SDSS sources per square degree, but when one cuts by stellarity
and magnitude this number drops to around 1000. These numbers are
valid at high galactic latitudes; for fields near the galactic plane,
the number of stars increases. It is best to only use the stars (the
above color terms are more appropriate to stars than galaxies), and to
only use the objects with $17<{\rm mag}<20$ (the brighter objects are
usually saturated in the MegaCam image and including the fainter
objects only increases the noise in the median). Still, with 1000
objects, the zero point relative to the SDSS calibration can be
determined to $0.07/\sqrt{1000}=\pm0.002$ magnitudes or better, even
for the \U\ band.  This process can be used any night; it is not
necessary for the night to be photometric.

For groups outside the SDSS footprint, the Elixir photometric keywords
are used, with modifications. The Elixir zero-points were compared to
those determined from the SDSS using the procedure above for a large
number of images. There are systematic offsets between the two sets of
zero-points, particularly for the \U-band. These offsets also show
variations with epoch, which are caused by modifications to Elixir
pipeline (Cuillandre, private communication).
The most significant effect is a 0.2 magnitude jump in the u-band
zero-point between Elixir data from before May 2006 and data
from after May 2006. 
For MegaPipe, the
offsets are applied to the Elixir zero-points to bring them in line
with the SDSS zero-points.

Archival data from the SkyProbe real-time sky-transparency monitor
\markcite{skyprobe}({Cuillandre} {et~al.} 2002) is used to determine if a night was photometric or
not.  Data taken on photometric nights is processed first through the
astrometric and photometric pipelines to generate a catalogue of
in-field standards. These standards are then used to calibrate any
non-photometric data in a group. If none of the exposures in a group
was taken on a photometric night, then that group cannot be processed.

\section{IMAGE STACKING}
\label{comb}
The calibrated images are coadded using the program \swarp\
\markcite{swarp}(Bertin 2004). \swarp\ removes the sky background from each image so
that its sky level is 0. It scales each image according to the
photometric calibration described in section \ref{photom}.  \swarp\
then resamples the pixels of each input image to remove the geometric
distortion measured in section \ref{astrom} and places them in the
output pixel grid which is an undistorted tangent plane projection.  A
``Lanczos-3'' interpolation kernel is used as discussed
in section 5.6.1 of Bertin (2004).  The values of the
flux-scaled, resampled pixels for each image are then combined into
the output image by taking a median. A median is noisier than an
average, but rejects images defects such as cosmic rays or bad columns
better.  The optimum would be some sort of sigma-clipped average, but
this is not yet an option in \swarp.

The input images are weighted with mask images provided as part of the
Elixir processing. The mask images have the value 1 for good data and
0 for pixels with known image defects. An inverse variance weight map
is produced along with each output image.
This can be used as an input
when running SExtractor on the stack.

The images are combined as full mosaics. The resulting output images
(stacks) measure about 20000 pixels by 20000 pixels or about 1 degree
by 1 degree (depending on the input dither pattern) and are about 1.7
Gb in size. They are scaled to have
a photometric zero-point of 30.000 in AB magnitudes; that is, for each
source:

\begin{equation}
{\rm AB magnitude} = -2.5 \times \log_{10}({\rm counts}) + 30.000.
\end{equation}

Due to the sky-substraction described above, the images have a sky
level of 0 counts.  This can cause unintended results for extended
objects.  If the extended emission varies smoothly and slightly (as in
a large nebula) SWarp's background subtraction removes the extended
emission at the same time as it removes the sky background, leaving a
blank field. If the extended emission has sharp variations (such as
the spiral arm of a nearby galaxy), SWarp's background subtraction can
produce peculiar results.

\section{CATALOGUE GENERATION}
\label{cat}
SExtractor \markcite{hihi}({Bertin} \& {Arnouts} 1996) is run on each output image stack using the
weight map. The resulting catalogues only pertain to a single band; no
multi-band catalogues have been generated. While this fairly simple
approach works well in many cases, it is probably not optimal in some
situations. SExtractor was originally designed for sparse, high
galactic latitude fields and does not do very well in highly crowded fields.
In these cases, some users may wish to run their own catalogue
generation software such as DAOphot \markcite{daophot}({Stetson} 1987).

\section{CHECKS ON ASTROMETRY}
\label{qualastro}
\subsection{Internal Accuracy}
The internal accuracy is checked by running SExtractor on each stacked
image in every band of each group and obtaining catalogues of object
positions. The positional catalogues for each band are matched to each
other and common sources are identified. If the astrometry is perfect,
then the position of the sources in each band will be identical. In
practice, there are astrometric residuals. Examining these residuals
gives an idea of the astrometric uncertainties.

\begin{figure}
\plotone{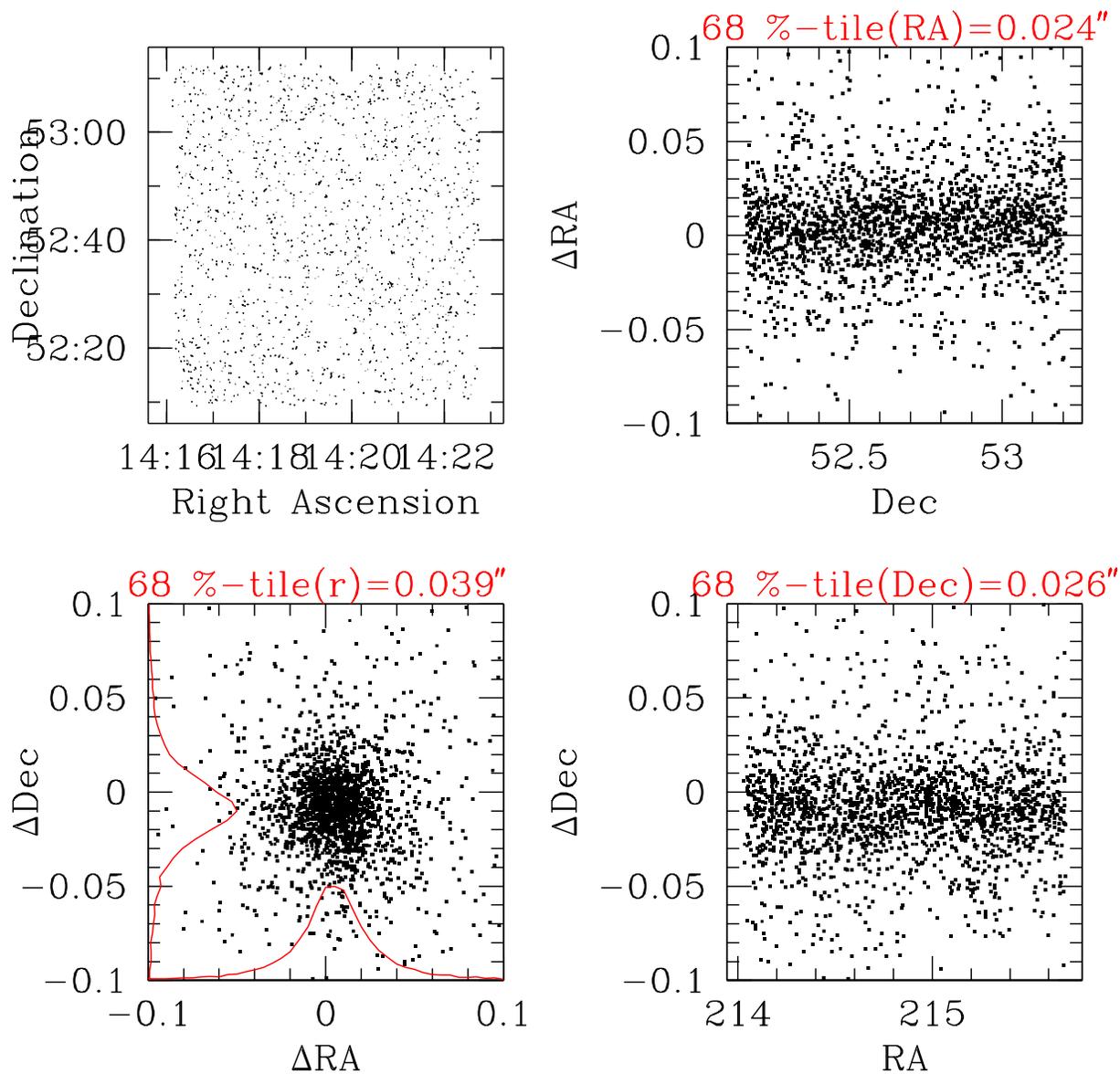}
\caption{An example of internal astrometric residuals.
The top left quarter shows the direction and size (greatly enlarged)
of the astrometric residuals as line segments.
The bottom left quarter shows the astrometric residuals in RA and
Dec. The red histograms show the relative distribution of the
residuals in both directions. 
The two right panels show the residuals in RA and Dec as functions
of Dec and RA respectively.
}
\label{fig:astint}
\end{figure}

\begin{figure}
\plotone{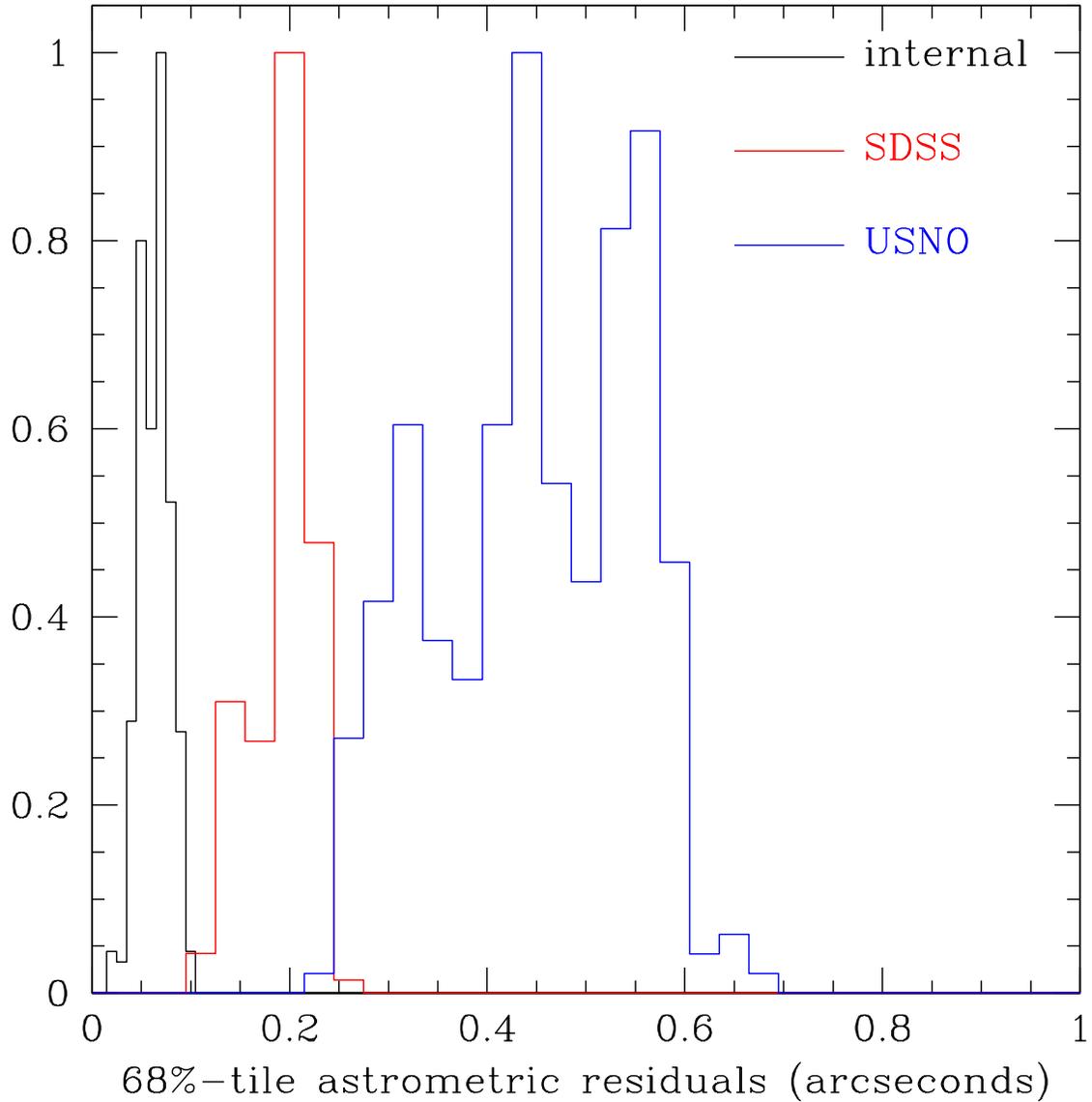}
\caption{The relative distribution of astrometric residuals for 300 groups.
The internal astrometric residuals (shown as a black line) are
typically 0.06 arcseconds.  The external astrometric residuals are
typically 0.2 with respect to the SDSS (shown as a red line) and 0.5 arcseconds
with respect to the USNO (shown as a blue line).  The SDSS is clearly a superior
astrometric reference.}
\label{fig:astrores}
\end{figure}

Figure \ref{fig:astint} shows checks on the internal astrometry
between two images in a group.  The top left quarter shows the
direction and size (greatly enlarged) of the astrometric residuals as
line segments. This plot is an important diagnostic of astrometry
because, while the residuals are typically quite small, there are
outliers in any distribution.  If these outliers are relatively
isolated from each other and pointing in random directions, this
indicates errors in cross-matching between the two images, and is not
an indicator of astrometric errors.  Conversely, if there are a number
of large residuals in close proximity to each other, all pointing in
the same direction, this indicates a systematic misalignment between
the two images in question. The figure shows no such misalignments.
The bottom left quarter of Figure \ref{fig:astint} shows the
astrometric residuals in RA and Dec. The red histograms show the
relative distribution of the residuals in both directions. The
68\%-tile of the residuals is 0.040 arcseconds radially.  The two
right panels show the residuals in RA and Dec as a function of Dec and
RA respectively. The error in each direction is about 0.025
arcseconds.  Note that there should be a factor of $\sqrt{2}$ between
the radial uncertainties and the single direction
uncertainties. Figure \ref{fig:astint} shows a better than average
case.  More typically, the astrometric residuals are 0.06 arcseconds,
as shown in Figure \ref{fig:astrores}.

\subsection{Repeatability}

The test described above is applied to every pair of images within
each of the groups with similar results. This of course might not be
too surprising, since the images were registered to each other in the
first place. For example, if there is \G\ and \I\ data in a group, the \I\ 
image is made first and then the \G\ data is astrometrically mapped to
the \I\ image as described in Section \ref{astrom}. Therefore, even if
there are systematic errors in the \I astrometry, the \G\ data is mapped
to the erroneous positions, and the residuals between the \G\ and \I\
image will still be small.

\begin{figure}
\plotone{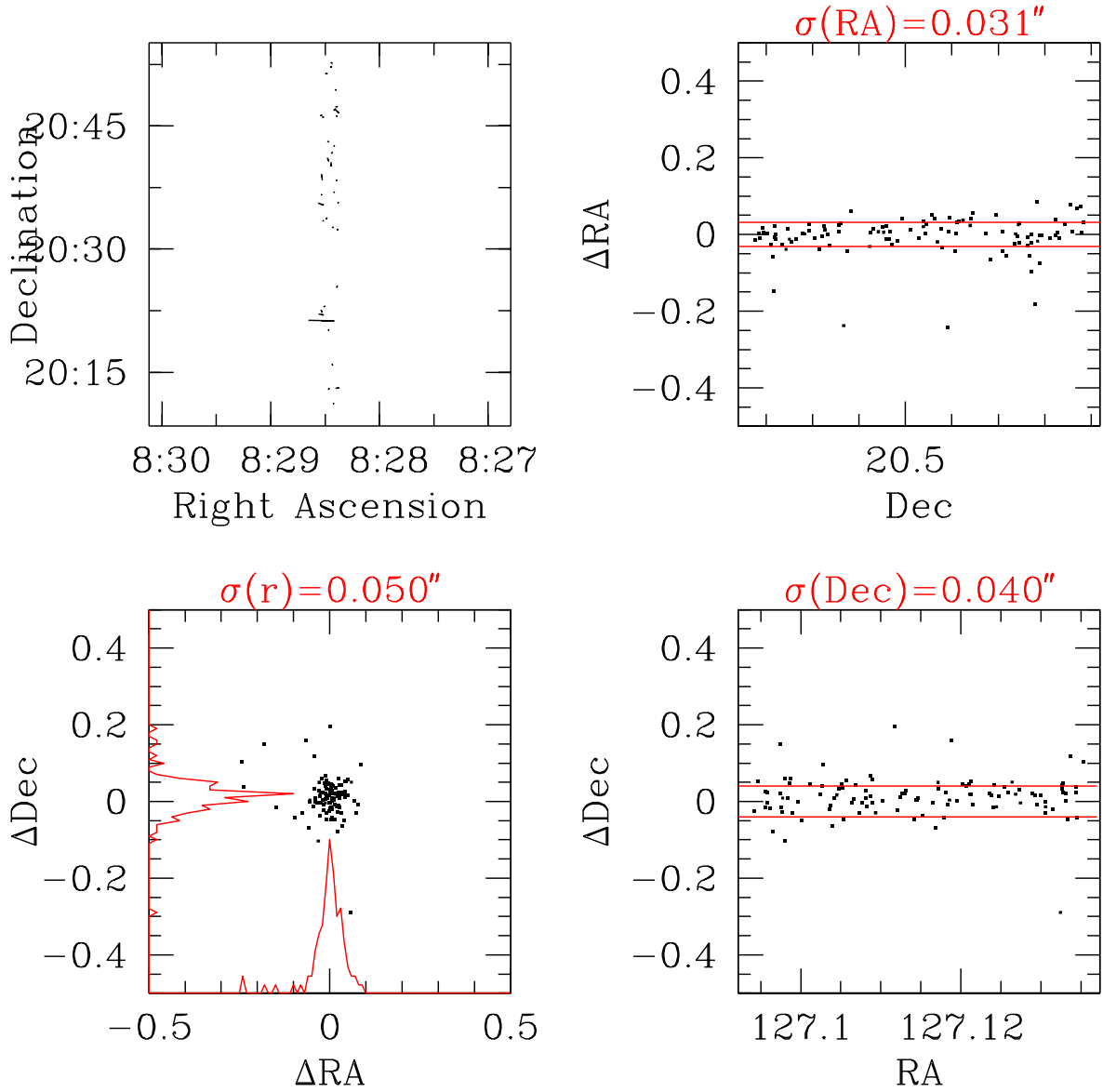}
\caption{Test of the astrometric repeatability. The panels have the same meaning
as in Figure \ref{fig:astint}.}
\label{fig:edgeast}
\end{figure}

However, this test is also applied between images belonging to
different groups. Since the astrometric calibration of one group is
completely independent to that of another group, comparing the
residuals between different groups is a more stringent test of the
repeatability of the astrometry.  Figure \ref{fig:edgeast} illustrates
this test. It shows the astrometric residuals between two groups. The
different panels have the same significance as in Figure
\ref{fig:astint}. Groups tend to overlap only at the edges, in a thin
strip, as shown in the top left panel. Consequently, the number of
common sources between two groups will be much smaller than between
two stacks in the same group. The residuals shown in Figure
\ref{fig:edgeast} case 0.05 arcseconds. More typically, averaging over
all the group overlaps, the repeatability is 0.06 arcseconds.

\subsection{External Accuracy}
\begin{figure}
\plotone{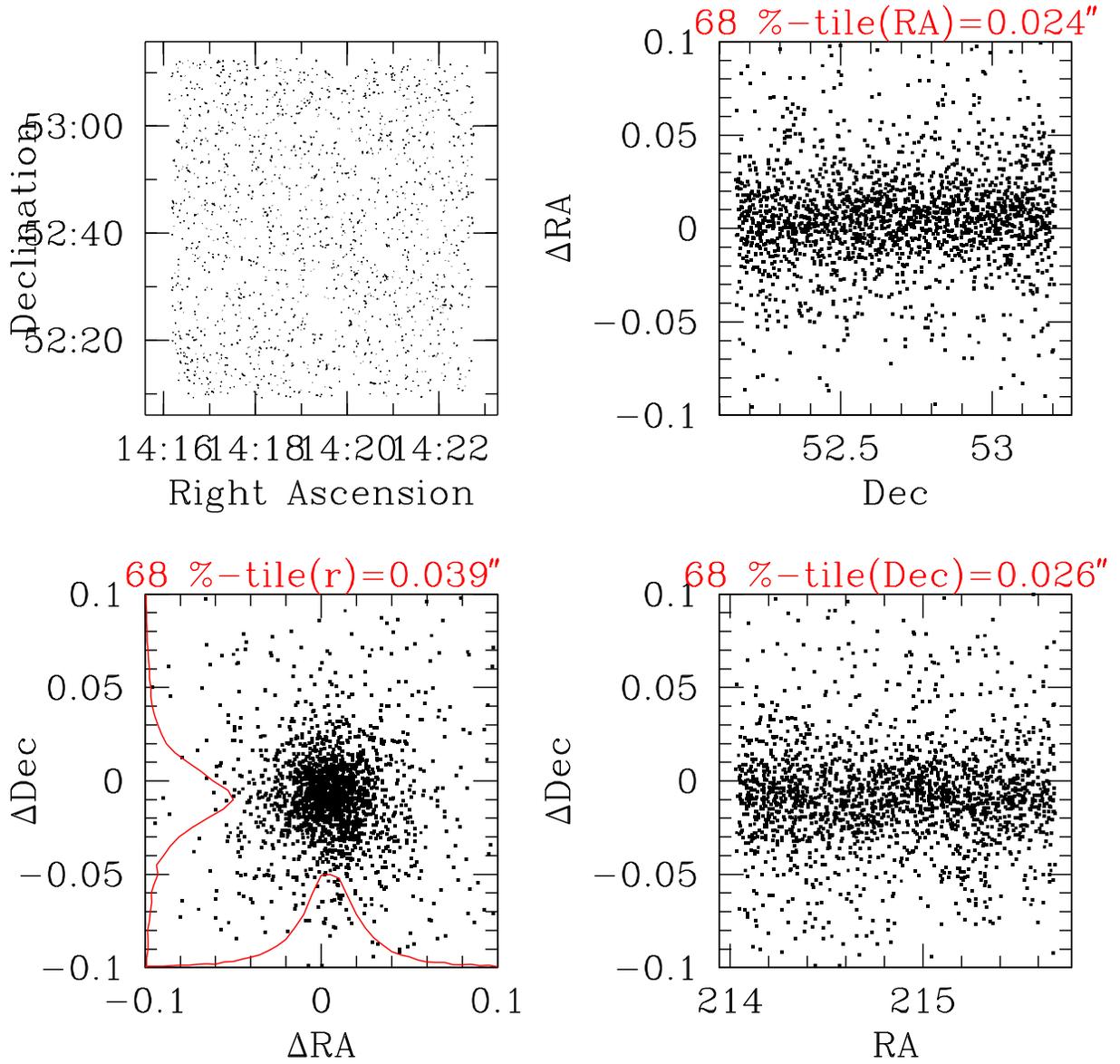}
\caption{An example of external astrometric residuals. The panels have the same meaning
as in Figure \ref{fig:astint}, but in this case the residuals are substantially larger.}
\label{fig:astext}
\end{figure}

This is checked by matching the catalogue for each field back to the
astrometric reference catalogue. Again, the scatter in the astrometric
residuals is a measure of the uncertainty, and the presence or absence
of any localized large residuals indicates a systematic shift.  Figure
\ref{fig:astext} shows checks on the external astrometry for a typical
group. The panels have the same significance as in Figure
\ref{fig:astint}. The only difference is that the residuals are
typically larger, generally in the neighborhood of 0.2 arcseconds.
Note that there are uncertainties in the external astrometric
catalogue as well. In this case, the SDSS is used as a reference. The
astrometric uncertainties inherent in the SDSS are 0.05 to 0.10
arceconds \markcite{pier2003}({Pier} {et~al.} 2003).  When this is taken into account, the
estimated external astrometric uncertainties are probably 0.15
arcseconds.  This is also true when the USNO is used as the external
catalogue.  The residuals are more typically 0.5, but the astrometric
uncertainties inherent in the USNO are about 0.3 or 0.4
arcseconds. The distribution of the astrometric residuals relative to
the USNO and SDSS for 300 different groups are shown in Figure
\ref{fig:astrores}.

\subsection{Image Quality}

Poor astrometry could also affect the stacked images in a more subtle
way. While internal astrometric errors are small (0.06 arcseconds)
compared to the pixels (0.18 arcseconds) and the typical image quality
(1 arcsecond or better due to the superb seeing conditions at CFHT),
they are not zero. Bad astrometry could degrade the image quality by
placing sources from the individual input images in slightly different
positions in the sky. When the images are coadded, the result would be
a smearing of the source, in effect increasing the seeing.

\begin{figure}
\plotone{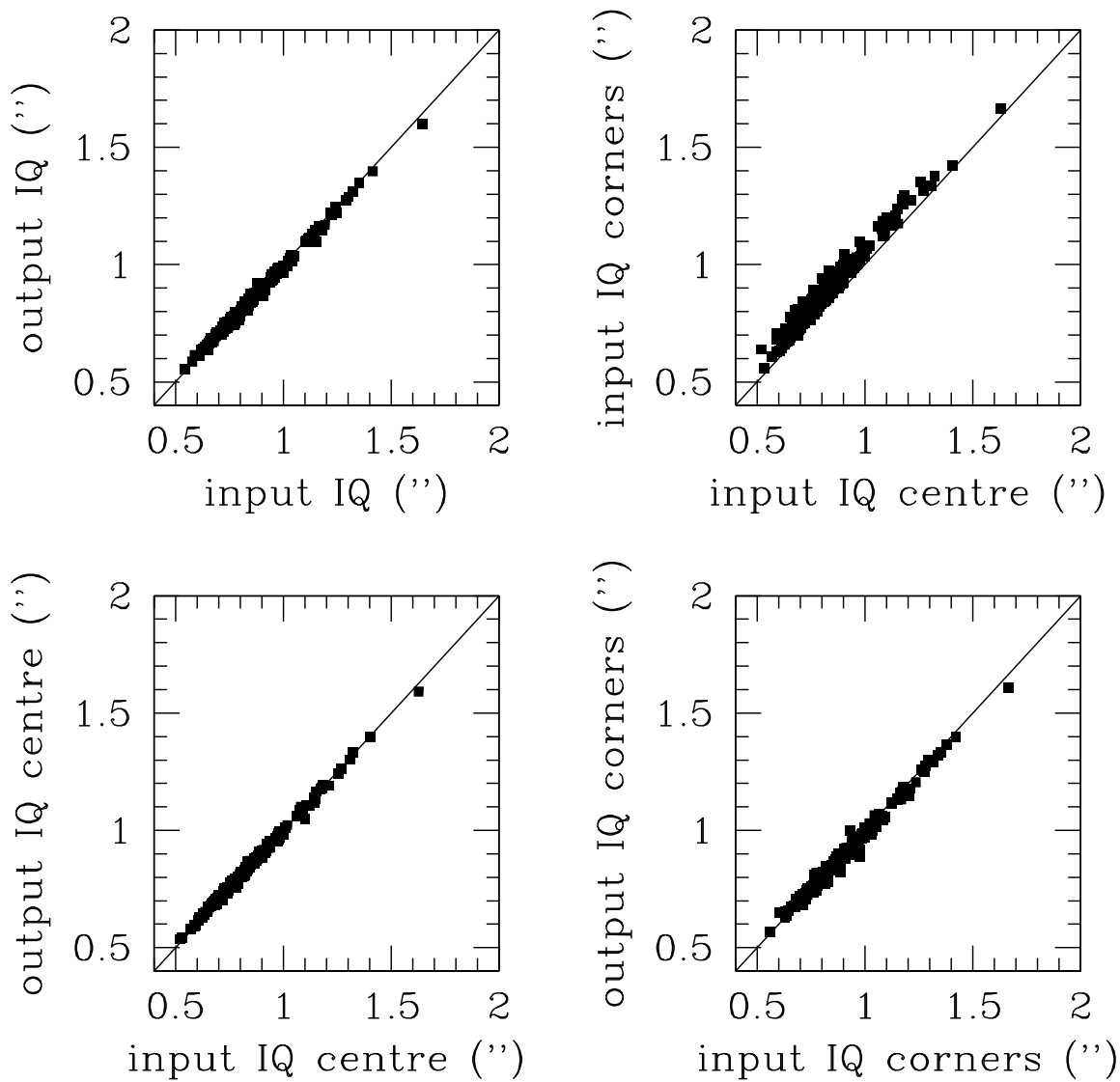}
\caption{Image quality. The top left plot shows the image quality (the ``seeing'') of the MegaPipe stacks
plotted against the median of the image quality of the input images that went into each stack.
The top right plot shows how the image quality degrades from the central portion of the mosiac
to the corners. The bottom two plots show that the image quality in both the centre and
the corners is not affected by the astrometric calibration or stacking process.}
\label{fig:iqinout}
\end{figure}

Figure \ref{fig:iqinout} shows that this is not taking place.  The top
left panel shows the image quality (the ``seeing'') of the MegaPipe
stacks plotted against the median of the image quality of the input
images that went into each stack. The figure shows that the output
image quality is the same as the median input image quality.  The
image quality degrades from the central portion of the mosiac to the
corners, as shown in the top right panel of Figure \ref{fig:iqinout}.
Here the, image quality has been determined separately for CCDs 00,
08, 27 and 35 of MegaCam (the four corners) and for CCDs 12, 13 14,
21, 22 and 23 (the centre).  The image quality is about 5\% worse at
the corners.  The bottom left and right plots show the effects of
stacking on the IQ for (respectively) the centre CCDs and the corner
CCDs. Again, the astrometric errors and the stacking process are not 
affecting the image quality.

\section{CHECKS ON PHOTOMETRY}
\label{qualphoto}

\subsection{Systematic Errors}
Whenever possible, the photometry of the MegaPipe images is directly
tied to the SDSS photometry. There $\sim1000$ standards in every
field. Thus, the systematic errors for these images are effectively
nil with respect to the SDSS. The systematic errors in the SDSS are
quoted as 2-3\% (Ivezic, et al., 2004).

The systematic errors for MegaPipe images not in the SDSS are limited
by the quality of the Elixir photometric calibration. 
Roughly half the images taken by MegaCam that can be calibrated lie
within the SDSS.  By comparing the Elixir photometric calibration to
the SDSS over a large number of nights, from 2003 until the present,
the systematic errors can be estimated.

The night-to-night
scatter is typically 0.02 to 0.03 magnitudes. Adding in quadrature the
SDSS systematic error (0.025 mags) to the systematic error in
transferring from the ``primary'' to ``secondary'' standards (0.025
mags), we get 0.035 magnitudes of total systematic error.

\subsection{External Comparisons}

For the groups lying within the SDSS, it is possible to check the
photometry of the stacked images directly. The magnitudes of sources
in the MegaPipe groups can be compared to the magnitudes from SDSS
transformed through the equations in Section \ref{photom}.  The
agreement is very good. The photometric differences can be entirely
attributed to the combination of the residuals in the equations
(\ref{eqn:trans1}) through (\ref{eqn:trans5}) and random errors.
The relative photometric offsets between the centre
of the mosaic and the corners is typically less than 0.005 magntiudes.

Of course, the comparison above applies to images that were
photometrically calibrated using the SDSS, so it is not surprising
that there are no residuals. As a test of the Elixir calibration, a
number of groups lying within the SDSS were stacked using only the
Elixir zero-points. The photometric residuals between the resulting
stacks and the SDSS were typically 0.03 magnitudes, consistent with
the photometric residuals between the individual (non-stacked) images
and the SDSS. 

\subsection{Internal Consistency}

The internal consistency is checked by comparing catalogues from
different groups. Groups occasionally overlap each other; even if they
only overlap by a arcminute or two, there are usually several hundred
sources in common in the two catalogues. Since groups are reduced
independently of each other, and since often the data was taken on
different nights, comparing the magnitudes of objects common to two
groups makes it possible to check the internal consistency.

\begin{figure}
\plotone{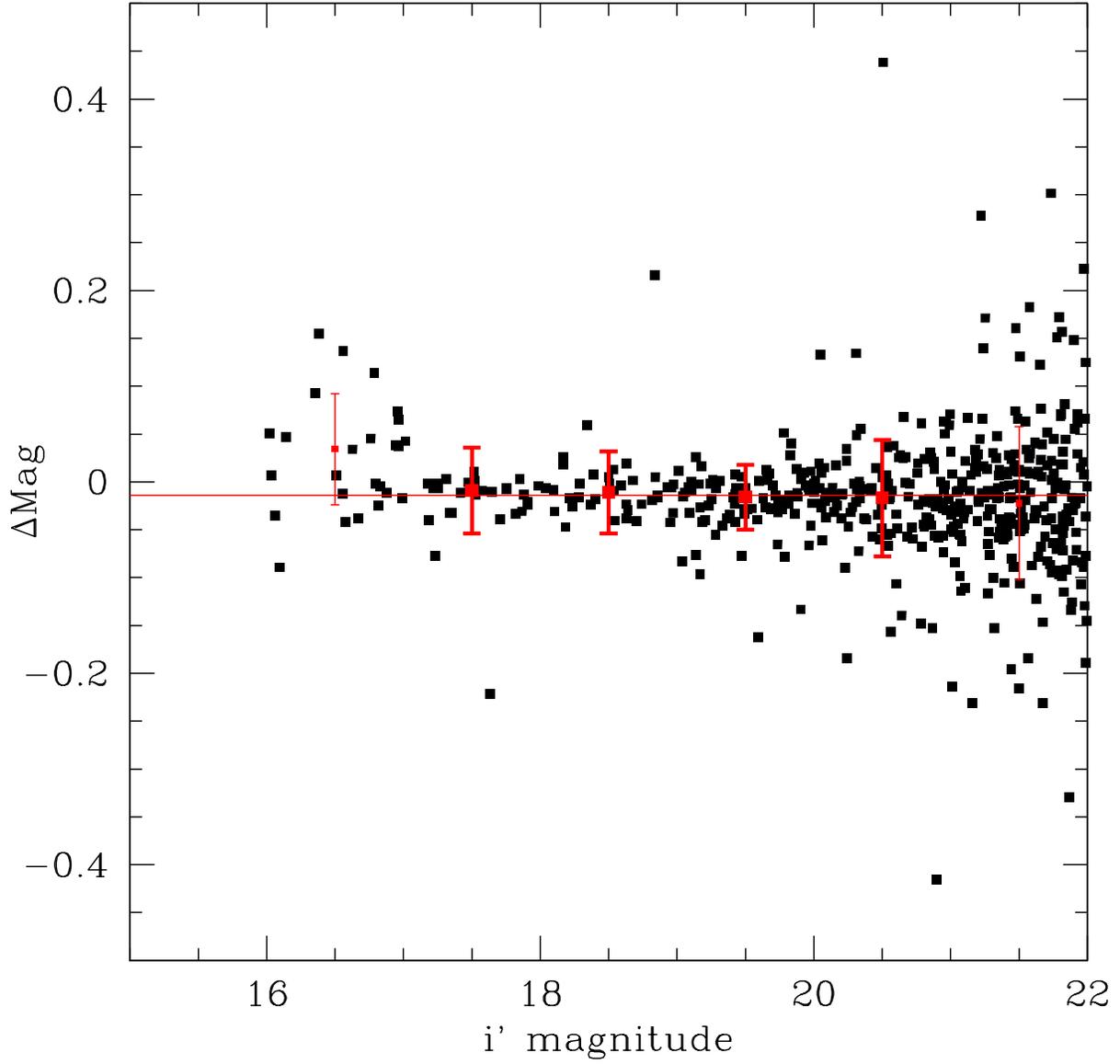}
\caption{A test of photometric repeatability. The photometry in two overlapping
groups is compared. The residuals are plotted against magnitude measured in one
of the groups. The points with error bars show median and 68\%-tile levels for
different magnitude bins.}
\label{fig:edgemag}
\end{figure}

Any systematic errors will show up as an offset in the median of the
difference in magnitudes between the two groups, as shown in Figure
\ref{fig:edgemag}. The offset in this case is -0.014 magnitudes. This
same test was applied to all possible pairs of groups where there were
more than 100 objects in common between the catalogues. The typical
offset was found to be 0.015 magnitudes. This is smaller than the
night-to-night variation of the Elixir zero-points (which is 0.03
mags) for two reasons: First, many groups lie within the SDSS so that their
photometric calibration does not depend on the Elixir
zero-points. Second, some neighbouring groups were observed on the
same night so any systematic error in the Elixir zero-points will be
common to both groups.

\subsection{Star Colors}

Another diagnostic of photometry is to examine the colors of
stars. Stars have a relatively constrained locus in color space. Any
offsets between the observed and expected colors indicates a zeropoint
error.  This test will of course only indicate failure if the shift is
quite large (0.05 magnitudes or more); smaller shifts are not
visible. Further, the metalicity of the star population will
systematically affect the color-color locus. High galactic latitude
fields (metal poor) will not look the same as fields on the galactic
plane (metal rich). Therefore, this test can not be viewed as
definitive.  However, the test can be applied to groups that do not
lie in the SDSS and therefore cannot be checked directly.

\begin{figure}
\plotone{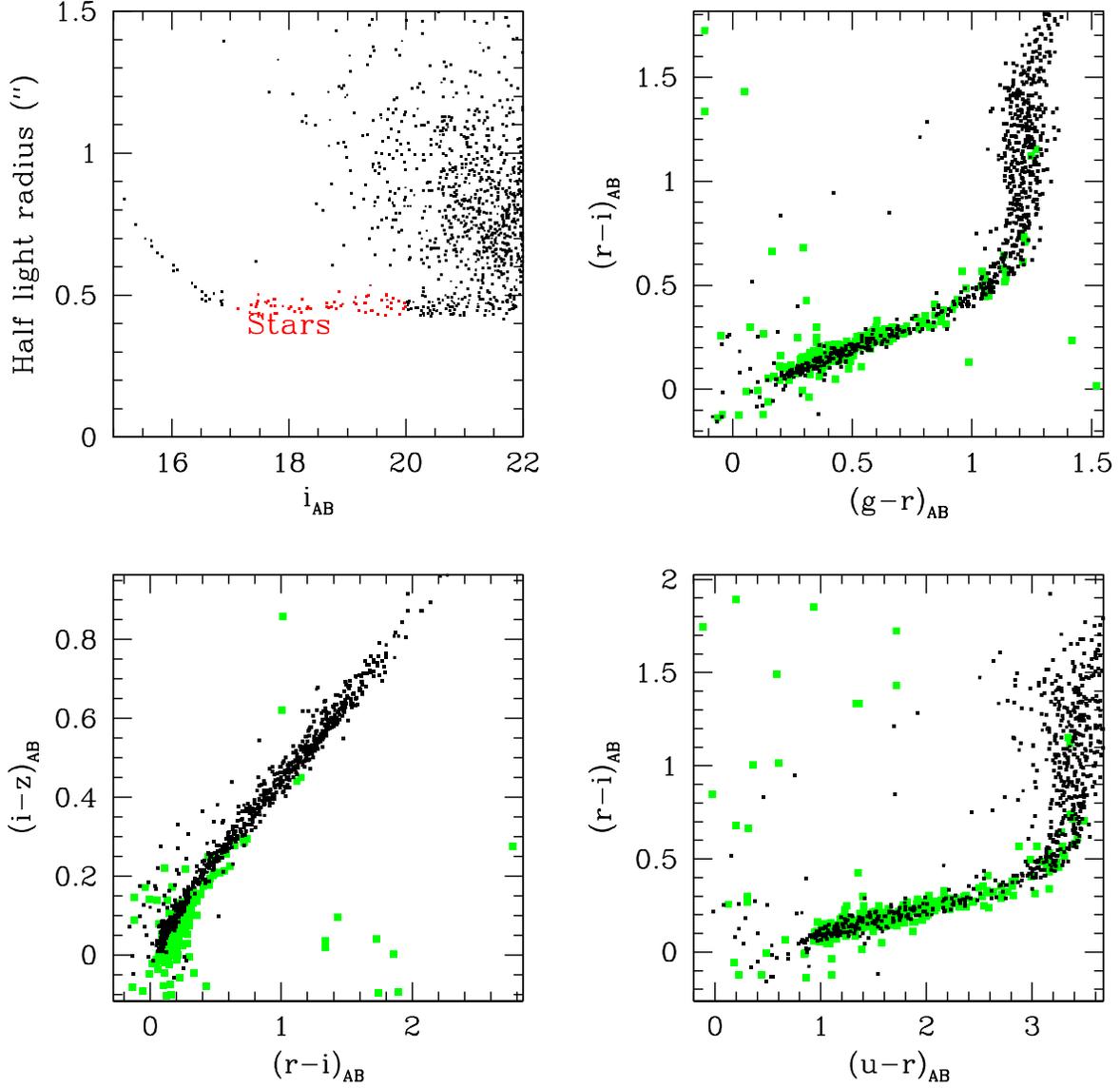}
\caption{Star colors.  The top left panel illustrates the selection of
stars for a typical image. The plot shows half-light radius plotted
against magnitude. On this plot, the galaxies occupy a range of
magnitudes and radii while the stars show up as a well defined
horizontal locus, turning up at the bright end where the stars
saturate. The red points indicate the very conservative cuts in
magnitude and radius to select stars for further analysis.  The other
3 panels show the colors of the stars selected in this manner in black
overlaid on top of the transformed SDSS star colors shown in green.  }
\label{fig:starcol}
\end{figure}

The top left panel of Figure \ref{fig:starcol} illustrates the
selection of stars for a typical image. The plot shows half-light
radius plotted as magnitude. On this plot, the galaxies occupy a range
of magnitudes and radii while the stars show up as a well defined
horizontal locus, turning up at the bright end where the stars
saturate. The red points indicate the very conservative cuts in
magnitude and radius to select stars for further analysis.

The other 3 plots show the colors of the stars selected in this manner
in black.  The underlying stellar locus (show in green) was generated
by selecting point sources from the SDSS and transforming the colors
by using Equations (\ref{eqn:trans1}) through (\ref{eqn:trans5}). No
systematic shifts seem to be visible.  The SDSS points which do not
lie on the stellar locus are quasars.  This test was applied for all
groups where stacks were made in 3 or more bands with similar results.

\subsection{Limiting Magnitudes}

The limiting magnitudes of the images is measured in three ways:
\begin{itemize}
    \item Number count histogram
    \item 5-sigma point source detection
    \item Adding fake objects
\end{itemize}

The first method is quite simple, indeed crude. The magnitudes of the
objects are sorted into a histogram. The peak value of the histogram,
where the number counts start to turn over, is a rough measure of the
limiting magnitude of the image.

The second method is also simple. The estimated magnitude error of
each source is plotted against its magnitude. In this case, the {\tt
MAG\underline{~}AUTO} or Kron-style \markcite{Kron}({Kron} 1980) magnitude is
plotted. At the faint magnitudes typical of MegaCam images, the sky
noise dominates over the magnitude error. This means that extended
objects (which have more sky in their larger Kron apertures) will be
noisier for a given magnitude than compact sources. Turning this
around, this means that, for a given fixed magnitude error, a point
source will be fainter than an extended source. A 5-sigma detection
corresponds to a S/N of 5 or, equivalently, a magnitude error of 0.198
magnitudes. Thus, to find the 5-sigma point source detection limit,
one finds the faintest source whose magnitude error is 0.198
magnitudes or less. It must be a point source, therefore, its
magnitude is the 5-sigma point source detection limit. A more refined
approach would be to isolate the point sources, by using the
half-light radius for example. In practice, the quick and dirty
method gives answers that are correct to within $\sim0.3$ magnitudes,
which is accurate enough for many purposes.

\begin{figure}
\plotone{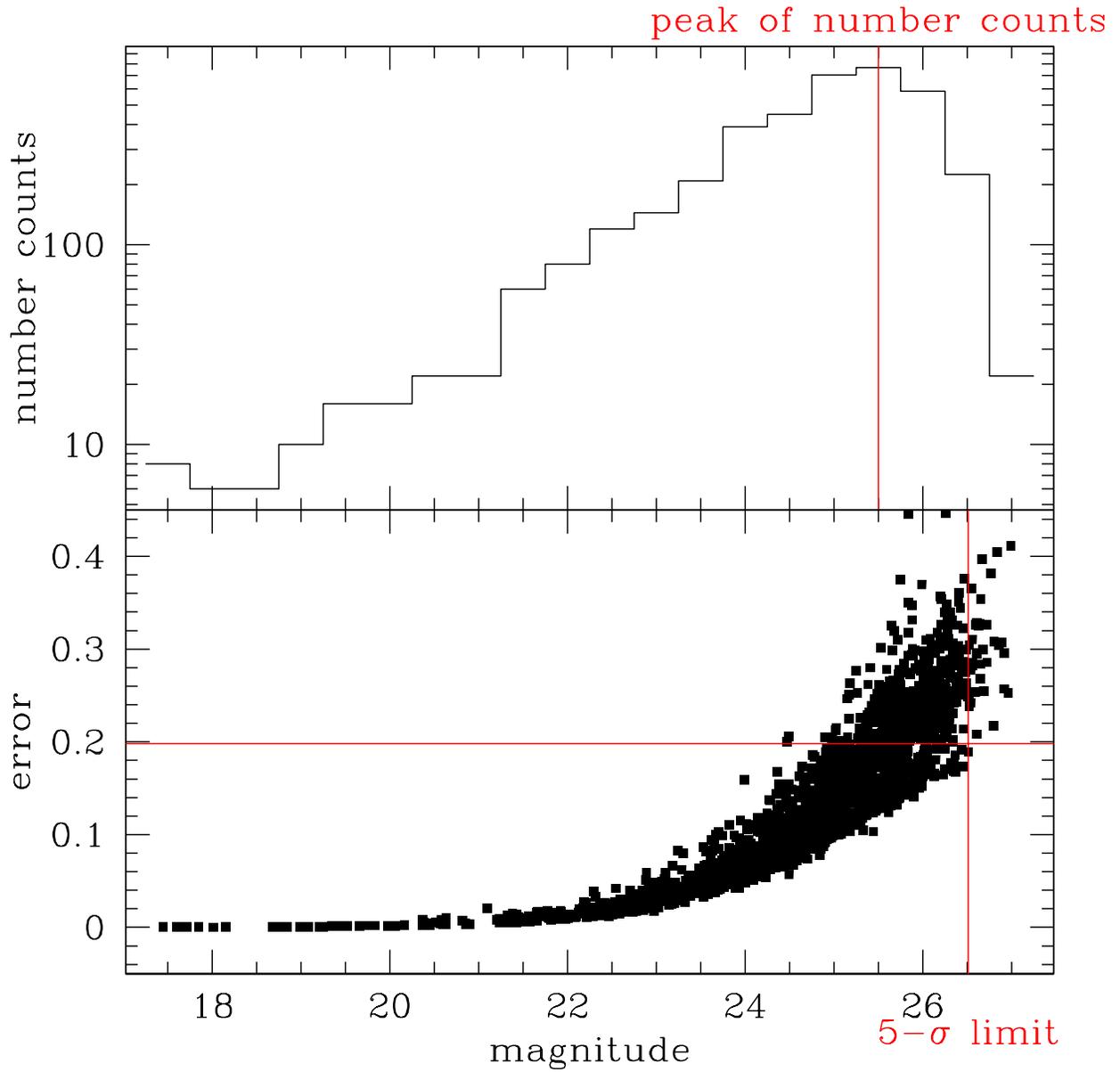}
\caption{Limiting magnitudes. The top panel illustrates the number count method
of measuring the limiting magnitude. The lower panel illustrates
the 5-$\sigma$ detection limit method.}
\label{fig:maglimnc}
\end{figure}

Figure \ref{fig:maglimnc} illustrates these methods. The top panel
shows the number count histogram. The number counts peak at 25.5 in
magnitude as shown by the vertical red line.

The bottom panel shows magnitude error plotted against magnitude. The
horizontal red line lies at 0.198 magnitudes. The vertical red line
intersects the horizontal line at the locus of the faintest object with
a magnitude error less than 0.198 magnitudes. The magnitude limit by
this method is 26.5 magnitudes.

\begin{figure}
\plotone{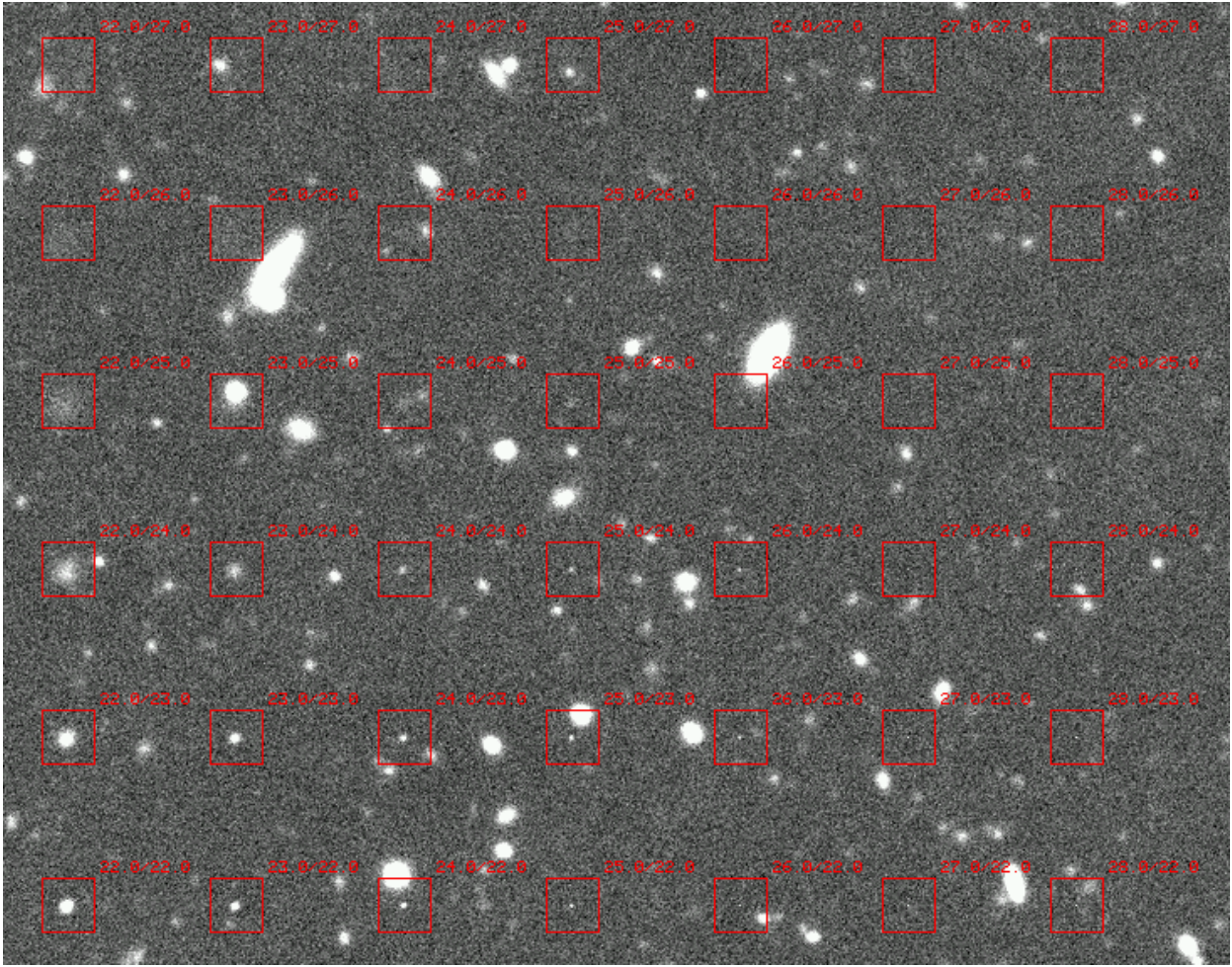}
\caption{Adding fake galaxies to an image. The same galaxy has been
added to the image repeatedly.  It has been artificially made fainter in
total magnitude/surface brightness as noted above and to the right of
the little red boxes.  Surface brightness decreases vertically and
total magnitude decreases horizontally.}
\label{fig:sampleim}
\end{figure}

The final way the limiting magnitudes of the images was tested by
adding fake galaxies to the images and then trying to recover them
using the same parameters used to create the real image
catalogues. The fake galaxies used are taken from the images
themselves, rather than adding completely artificial galaxies. 40
bright, isolated galaxies are selected out of the field. Postage
stamps of these galaxies are cut out of the images. The galaxies are
faded in both surface brightness and magnitude through a combination
of scaling the pixel values and resampling the images.  To test the
recovery rate at a given magnitude and surface brightness, galaxy
postage stamps are selected from the master list, faded as described
above to the magnitude and surface brightness in question, and then
added back to the image at random locations. SExtractor is then run on
the new image. The fraction of fake galaxies found gives the recovery
rate at that magnitude and surface brightness. An illustration of
adding fake galaxies is shown in Figure \ref{fig:sampleim}. The same
galaxy has been added multiple times to the image, faded to various
magnitudes and surface brightnesses. The red boxes contain the galaxy
and are labelled by magnitude/surface brightness. Note the galaxy at $\I=23$,
$\mu_{\I}=25$ accidentally ended up near a bright galaxy and is only
partially visible. Normally of course, the galaxies are not placed in
such a regular grid.

To test the false-positive rate, the original image was multiplied by
-1; the noise peaks became noise troughs and vice-versa. SExtractor
was run, using the same detection criteria. Since there are no real
negative galaxies, all the objects thus detected are spurious.

\begin{figure}
\plotone{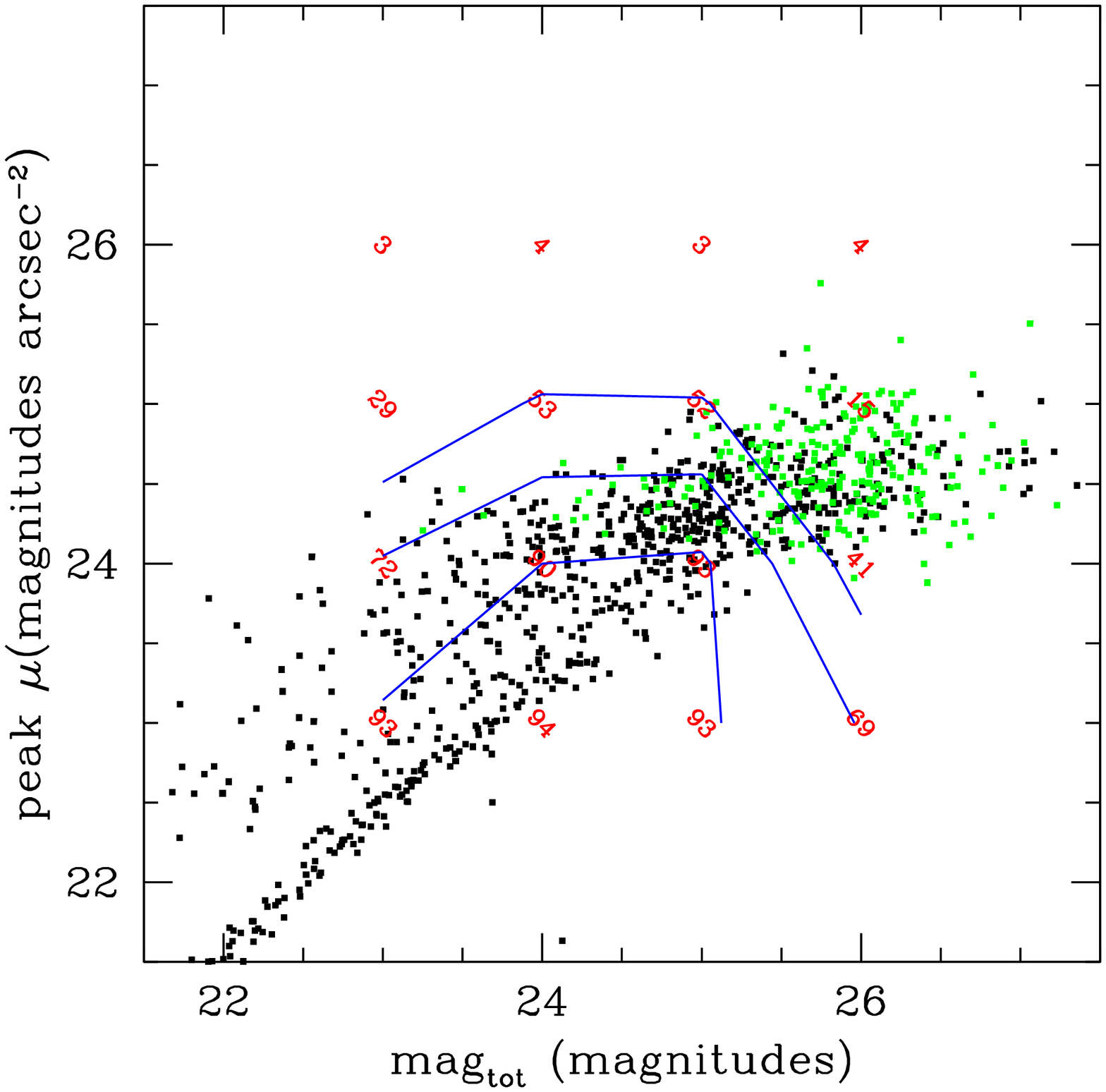}
\caption{Limiting magnitude and surface brightness.
The black points are real
objects. The bottom edge of the black points is the locus of
point-like objects. The green points show the false-positive
detections. The red numbers show the percent of artificial galaxies
that were recovered at that magnitude/surface brightness. The blue
contour lines shows the 90\%, 70\% and 50\% completeness levels.  
Point sources occupy a distinct, diagonal locus slightly below the
$\mu={\rm mag}_{\rm tot}$ line. Extended objects lie above this.
}
\label{fig:maglimmu}
\end{figure}

The magnitude/surface brightness plot shown in Figure
\ref{fig:maglimmu} is an example of such a test. The black points are
real objects. The bottom edge of the black points is the locus of
point-like objects. The green points show the false-positive
detections. The red numbers show the percentage of artificial galaxies
that were recovered at that magnitude/surface brightness. The blue
contour lines show the 90\%, 70\% and 50\% completeness levels.

Deriving a single limiting magnitude from such a plot is slightly
difficult. The cleaner cut in the false positives seems to be in
surface brightness. Extended objects become harder to detect at
brighter magnitudes whereas stellar objects are detectable a magnitude
or so fainter.

Note that this plot is of limited usefulness in crowded fields. In
this case, an object may be missed even if it is relatively bright
because it lies on top of another object. However, the objects in crowded
fields are almost always stellar. This suggests the use of the DAOphot
package rather than using the SExtractor catalogues provided as part of
MegaPipe.

\section{PRODUCTION AND DISTRIBUTION}
\label{sec:prod}
The MegaPipe pipeline is now in place at the Canadian Astronomical
Data Centre\footnote{\url{http://www.cadc.hia.nrc.gc.ca/megapipe/}}.
The rate at which stacks can be generated depends directly on the
number of input images.  With the current generation of processing
nodes at the CADC, each group can be produced in 10 minutes $\times$
the number of input images. This last number is included chiefly to
amuse future generations of astronomers.

At present, over 700 groups have been generated with a total of about
1500 stacks comprising 12000 input images.  The plan is to process all
MegaCam images as they become public.

The images are distributed via the Canadian Astronomical Data Centre.
A user can search for images by position or by name, or by the
properties of the input images (number of input images, total exposure
time, etc.). A preview facility is provided which allows the user to
rapidly pan and zoom over the images without downloading the fairly
sizable science images. A cutout service which allows users to
retrieve a small subsection of a MegaPipe image is also provided.

\acknowledgments

This research used the facilities of the Canadian Astronomy Data
Centre operated by the National Research Council of Canada with the
support of the Canadian Space Agency.

S.D.J.G is an NSERC Visiting Fellow in a Canadian Government Laboratory.

Based on observations obtained with MegaPrime/MegaCam, a joint project
of CFHT and CEA/DAPNIA, at the Canada-France-Hawaii Telescope (CFHT)
which is operated by the National Research Council (NRC) of Canada,
the Institute National des Sciences de l'Univers of the Centre
National de la Recherche Scientifique of France, and the University of
Hawaii.

{\it Facilities:} \facility{CFHT}.


\clearpage

\clearpage

\clearpage

\clearpage

\clearpage

\clearpage

\clearpage

\clearpage

\clearpage

\clearpage

\clearpage

\clearpage

\end{document}